\documentclass[
 reprint,
superscriptaddress,
 amsmath,amssymb,
 aps,
]{revtex4-2}

\usepackage{graphicx}
\usepackage{dcolumn}
\usepackage{bm}
\usepackage{upgreek,xspace}
\usepackage[linesnumbered,ruled,lined]{algorithm2e}
\usepackage{enumitem}   

\usepackage{hyperref}
\hypersetup{colorlinks, linkcolor={red},citecolor={blue}, urlcolor={black}}

\begin{document}

\title{Metasurface-Programmable Wireless Network-on-Chip}

\author{Mohammadreza F. Imani}
 \affiliation{School of Electrical, Computer, and Energy Engineering, Arizona State University, Tempe, Arizona 85287, USA}
 
 \author{Sergi Abadal}
 \affiliation{NaNoNetworking Center in Catalunya (N3Cat), Universitat Politècnica de Catalunya, 08034
Barcelona, Spain}

 \author{Philipp del Hougne}
 \email{philipp.del-hougne@univ-rennes1.fr}
 \affiliation{Univ Rennes, CNRS, IETR - UMR 6164, F-35000, Rennes, France}

\begin{abstract}
We introduce the concept of smart radio environments, currently intensely studied for wireless communication in metasurface-programmable meter-scaled environments (e.g., inside rooms), on the chip scale. Wireless networks-on-chips (WNoCs) are a candidate technology to improve inter-core communication on chips but current proposals are plagued by a dilemma: either the received signal is weak, or it is significantly reverberated such that the on-off-keying modulation speed must be throttled. Here, we overcome this vexing problem by endowing the wireless on-chip environment with \textit{in situ} programmability which allows us to shape the channel impulse response (CIR); thereby, we can impose a pulse-like CIR shape despite strong multipath propagation and without entailing a reduced received signal strength. First, we design and characterize a programmable metasurface suitable for integration in the on-chip environment (``on-chip reconfigurable intelligent surface''). Second, we optimize its configuration to equalize selected wireless on-chip channels ``over the air''. Third, by conducting a rigorous communication analysis, we evidence the feasibility of significantly higher modulation speeds with shaped CIRs. Our results introduce a programmability paradigm to WNoCs which boosts their competitiveness as complementary on-chip interconnect solution.

\end{abstract}

\maketitle

\section{Introduction}
\label{sec:intro}

\begin{figure*}
    \centering
    \includegraphics[width=1\linewidth]{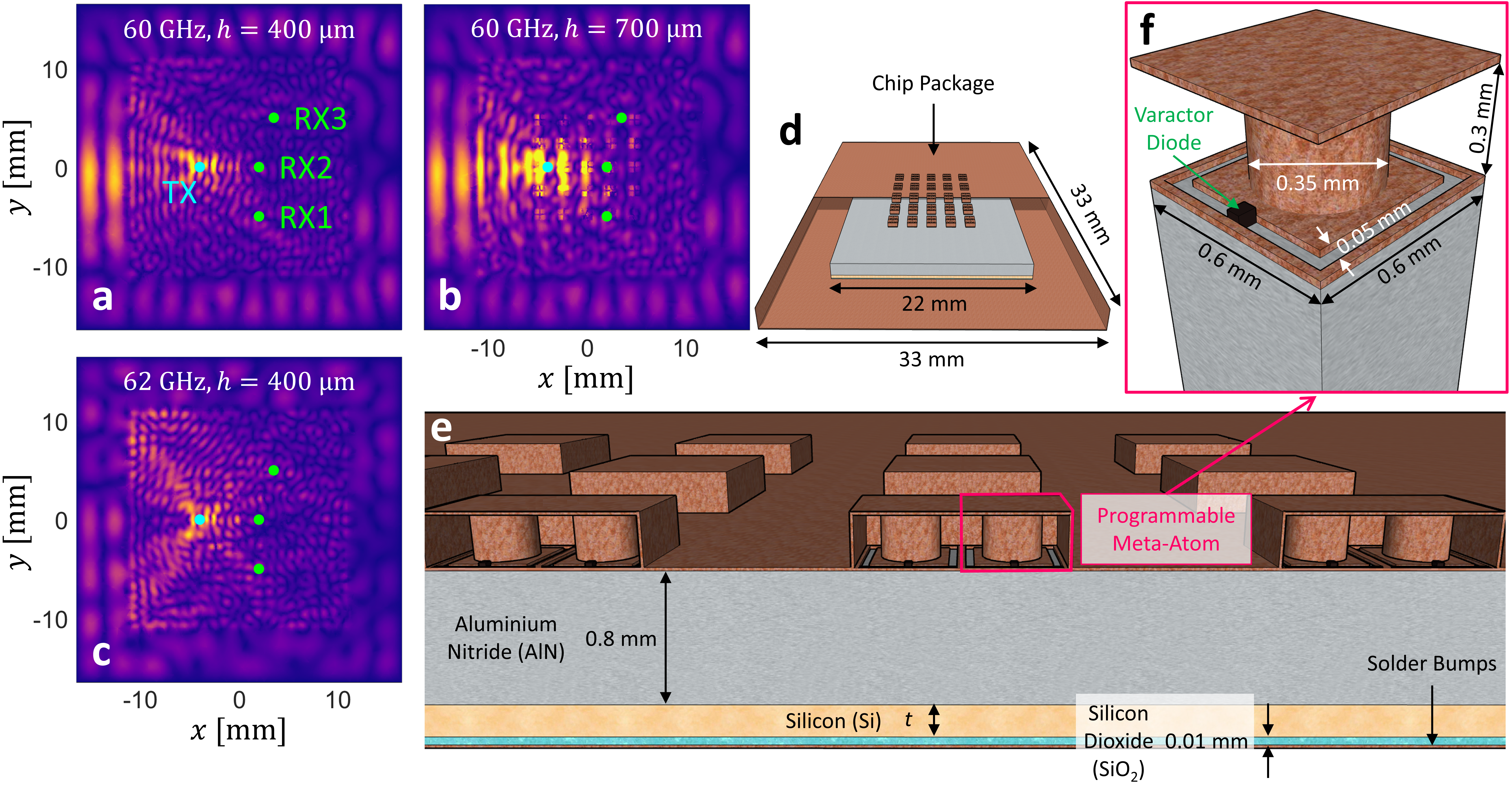}
    \caption{\textbf{Metasurface-programmable wireless on-chip environment.}
    (a-c) Maps of the spatial field magnitude (for the field component that is perpendicular to the chip plane) along a horizontal slice inside the chip package shown in (d,e) with $t=100\ \upmu\mathrm{m}$ for the indicated parameters (frequency and vertical height $h$). The colormap is linear. In addition, the transmitter (TX) and receiver (RX) locations considered in this paper are indicated in (a).
    (d) Overview of the considered chip architecture. The chip is equipped with a $5\times 5$ programmable metasurface in its ceiling. A part of the chip package is removed in this figure to show the interior.
    (e) Vertical slice through the middle of the chip shown in (d), revealing the different layers and the programmable meta-atoms.
    (f) Detailed design of the considered programmable meta-atom.}
    \label{fig:general}
\end{figure*}

Wireless millimeter-wave (mmW) communication between processors on multi-core chips is a potential solution to avoid that inter-core information exchange soon becomes a computation speed bottleneck~\cite{deb2012wireless,bertozzi2015fast,rayess2017antennas,zhao2008sd}. Yet, such a wireless network-on-chip (WNoC) is confronted with its own challenges: either the received signals are too weak, or severe multipath curbs the information transfer rate~\cite{abadal2019wave}. In this Article, we overcome this dilemma by endowing the on-chip electomagnetic (EM) propagation environment with programmability. To that end, we integrate a programmable metasurface~\cite{sievenpiper2003two,holloway2005reflection,cui2014coding}, also referred to as reconfigurable intelligent surface (RIS), into the chip package (see Fig.~\ref{fig:general}d,e) -- analogous to current RIS-based efforts at the indoor scale~\cite{subrt2012intelligent,liaskos2018new,di2019smart,wu2019towards,basar2019wireless,RichScatteringRIS_magaz}. We demonstrate that the on-chip RIS can be configured such that the channel impulse response (CIR) becomes pulse-like despite strong multi-path propagation and without yielding weak received signals. We further evidence that shaped CIRs can sustain significantly larger modulation speeds and still satisfy a given bit-error-rate (BER) objective.

Traditionally, wireless communication is optimized in terms of transceiver hardware and pre-/post-processing of the signals but the propagation medium is considered uncontrolled. Recently, the emergence of programmable metasurfaces -- ultrathin arrays of elements with individually reconfigurable scattering response -- has led to a paradigm shift: ``smart'' programmable wireless propagation environments~\cite{subrt2012intelligent}. 
Therein, programmable metasurfaces are leveraged as RISs to shape the wireless channels. The desired channel-shaping functionality largely depends on the amount of scattering in the environment: 
\begin{itemize}
  \item In \textit{free space}, RISs, in combination with carefully aligned emitters, can replace costly phased-array antennas for beam-forming~\cite{cui2014coding,dai2020reconfigurable,shlezinger2019dynamic,tang2020mimo}.
  \item In \textit{quasi-free space} with blocked line-of-sight, RISs can serve as alternative relaying mechanism~\cite{hu2018beyond,huang2019reconfigurable,tang2020wireless,arun2020rfocus,di2020reconfigurable}.
  \item In \textit{rich-scattering} environments, wireless propagation is qualitatively sharply different from the previous two cases~\cite{RichScatteringRIS_magaz}. Rich scattering occurs, for example, inside irregularly shaped metallic enclosures such as vessels, air planes, trains, or cars, as well as inside certain indoor settings in buildings~\cite{Kaina2014SciRep,del2019optimally}. In rich-scattering scenarios, the field at any given point is a seemingly random superposition of waves arriving from all possible angles with all possible polarizations and diverse delays. Consequently, the field pattern is speckle-like~\cite{simon2001communication} and direct line-of-sight links are insignficant or inexistent. Such rich-scattering conditions are qualitatively very different from multipath scenarios solely involving a few \textit{known} scatterers. In rich-scattering environments, RISs can create spatial monochromatic hotspots~\cite{Kaina2014SciRep,dupre2015wave,del2016intensity,del2017shaping}, shape the multipath CIR~\cite{del2016spatiotemporal,RichScatteringRIS_magaz}, or optimize the spatial MIMO channel diversity~\cite{del2019optimally}.
\end{itemize}
There are also cases of backscatter communication in which the RIS encodes information \textit{by} shaping wireless channels~\cite{zhao2020metasurface,imani2020perfect}. To date, all of these ideas are explored for meter-scaled environments, such as office rooms. 

\clearpage
Two to three orders of magnitude smaller are on-chip wireless environments. The reason to consider a partial~\cite{franques2021widir} replacement of conventional wired interconnects on chips with WNoCs are scalability limits: more and more processing cores are crammed onto modern chips but wired interconnects must be kept short due to Ohmic losses and wire delays, leading to more and more inter-router hops. Consequently, both latency and power consumption of communication between far-apart cores deteriorate~\cite{marculescu2008outstanding}, the latency reaching up to several tens of nanoseconds~\cite{wentzlaff2007chip}. As a result, communication rather than computation is becoming the main performance bottleneck of multicore chips. To break these communication-related scalability barriers of current multicore architectures, complementary interconnect technologies are currently being explored. On the one hand, these involve approaches based on guided waves such as nanophotonic networks~\cite{miller2009device,dionne2010silicon,kurian2010atac,batten2013designing,sun2015single} or radio-frequency transmission lines~\cite{carpenter2011case,oh2013traffic} which benefit from energy efficiency and a large bandwidth; however, guided approaches intrinsically rely on physical infrastructure to connect the nodes which scales unfavorably because it requires increasingly stronger sources or more amplifiers, centralized arbitration, etc. On the other hand, WNoCs~\cite{chang2001rf,matolak2012wireless} avoid inter-router hops and promise low-latency broadcasting combined with intrinsic system-level flexibility. 

However, on-chip antennas and radio transceivers~\cite{floyd2002intra,kim2005chip,zhang2007propagation,lin2007communication,lee2009low,gutierrez2009chip,cheema2013last,yu20141,yu2014architecture,TSV,markish2015chip,wu2017monopoles} are subject to size and power constraints; in particular, their processing power is limited such that WNoCs typically rely on simple modulation schemes such as on/off keying (OOK)~\cite{laha2014new,markish2015chip}. 
Thus, inter-symbol interference (ISI) must be avoided in WNoCs at the cost of lower data transmission rates when the CIR is lengthy due to multipath. From the EM wave's perspective, a typical on-chip environment constitutes a metallic enclosure (solder bumps on the bottom, metallic package on sides and top) -- a ``micro reverberation chamber''~\cite{matolak2013channel}. While this enclosure seals the EM on-chip environment from the outside, making it extraordinarily static, predictable and secure, the enclosure also causes waves to heavily reverberate, yielding lengthy CIRs and the associated ISI problem. Reverberation can be suppressed through strong attenuation of the waves (e.g., by a thick silicon layer). But strong attenuation implies poor received-signal-strength-indicators (RSSIs). This yields the on-chip RSSI-ISI dilemma: either we face poor RSSI or the ISI problem. Poor RSSIs can be counteracted by transmitting stronger signals~\cite{mineo2015runtime} and trade-offs between the RSSI and ISI effects can be found~\cite{timoneda2020engineer}, but ideally one would have high-RSSI channels with pulse-like CIRs \textit{despite} strong multipath. 

In this Article, we demonstrate that this goal is potentially attainable if the CIR can be shaped with an on-chip RIS. In other words, we show that a RIS-parametrized WNoC can be programmed to equalize on-chip wireless channels ``over-the-air''. From an information theoretic perspective, it is important to note that \textit{in general} a pulse-like CIR is \textit{not} optimal in terms of the channel capacity, which is a modulation-independent upper bound on the information transfer rate~\cite{Shannon,Foschini,Telatar}. Indeed, rich-scattering-induced multipath effects can boost the achievable capacity by making different channels more  distinguishable~\cite{moustakas2000communication,simon2001communication}, but suitable equalization protocols must be applied to the received signals. Nonetheless, for the \textit{specific} simple modulation scheme used in WNoC, namely OOK, one can faithfully expect that the achievable information transfer is most reliable and efficient with a pulse-like CIR. Inside the static chip enclosure, there is hence a clear motivation to mitigate ISI by using an on-chip RIS to impose a pulse-like shape of the CIR. In Sec.~\ref{sec:communication} of this Article, we evidence that the bit-error-rate (BER, a modulation-specific metric) with OOK can indeed be orders of magnitude lower for a given modulation speed and noise level if an on-chip RIS is configured to impose a pulse-like CIR. 

The concept of RIS-empowered CIR shaping is also relevant to other wireless communication settings which  involve transceivers with low computational power facing lengthy CIRs, e.g., Internet-of-Things devices in indoor environments. A rigorous demonstration of the concept was experimentally reported at the indoor scale inside a rich-scattering enclosure for the 2.5~GHz range in Ref.~\cite{del2016spatiotemporal}. 
The fundamental underlying physical mechanism is based on altering the delays of the different multipath rays such that they interfere constructively (destructively) at (before/after) the time of arrival of the main CIR tap~\cite{del2016spatiotemporal}. This mechanism is hence \textit{not} based on the absorption of all multipath components. 
More recently, the concept of RIS-empowered ``over-the-air~equalization'' has also been explored based on channel models~\cite{zhang2021spatial,basar_frontiers,arslan2021over,zhou2021modeling,PhysFad}. However, Refs.~\cite{basar_frontiers,zhou2021modeling} consider a simple free-space two-path channel, and the remaining works,
with the exception of Ref.~\cite{PhysFad}, utilize channel models that are not compatible with all aspects of wave physics. Specifically, they appear to violate causality, to neglect the long-range mescoscopic correlations at the origin of the \textit{non}-linear RIS-parametrization of the fading wireless channels~\footnote{Experiments in rich-scattering wireless environments clearly show that the impact of a given RIS element on the wireless channel is \textit{not} independent of the other RIS elements' configuration (see, e.g., Refs.~\cite{dupre2015wave,del2016spatiotemporal}). The reason for the RIS-parametrization's non-linear nature are reverberation-induced long-range mesoscopic correlations.}, and to ignore the frequency selectivity of the RIS element as well as the intertwinement of its amplitude and phase response. 
To ensure the reliability of the present work, we use full-wave simulations which are computationally costly but guaranteed to respect wave physics.

The remainder of this paper is organized as follows. In Sec.~\ref{sec:dilemma}, we summarize a prototypical chip architecture and illustrate the ensuing RSSI-ISI dilemma. In  Sec.~\ref{sec:design}, we detail our design considerations for an on-chip RIS and thoroughly characterize our proposed design. In  Sec.~\ref{sec:shaping}, we demonstrate on-chip RIS-empowered CIR shaping. In  Sec.~\ref{sec:communication}, we evaluate the associated advantages for the achievable BER. 
In Sec.~\ref{sec:discussion}, we discuss our results. In Sec.~\ref{sec:conclusion}, we summarize our contributions and identify avenues for future research. Details on our methods are provided in Sec.~\ref{sec:methods}.

\section{On-Chip RSSI-ISI Dilemma}
\label{sec:dilemma}

To elucidate the RSSI-ISI dilemma, we analyze the impact of the silicon layer's thickness on RSSI and ISI. We rely on full-wave simulations (detailed in the Methods Sec.~\ref{subsec:cst}) of the simplified model of the on-chip wireless environment that is shown in Fig.~\ref{fig:general}. We consider a prototypical $22\times 22\ \mathrm{mm}^2$ chip inside a metallic package of footprint $33\times 33\ \mathrm{mm}^2$, as shown in in Fig.~\ref{fig:general}d; such a chip could host $4\times 4$ cores surrounded by air. The chip, illustrated in Fig.~\ref{fig:general}e, is a layered structure consisting of a very thin layer of solder bumps (a typical thickness is $0.0875\ \mathrm{mm}$) on top of the package substrate, followed by a 0.011~mm thick silicon-dioxide ($\text{SiO}_2$) layer, a
silicon (Si) substrate layer of thickness $t$ and finally a 0.8~mm thick aluminium nitride (AlN) layer~\cite{imani2020toward}. For simplicity, we assume that these layers are continuous without physical gaps between different cores. Moreover, we consider electrically small slot antennas as ports (see Methods Sec.~\ref{subsec:cst}). 
These almost omni-directional antennas have a low directivity, in line with most antennas proposed for WNoCs, because WNoC antennas must broadcast information to receivers located in all possible directions around them.

The rich-scattering nature of the wireless on-chip environment is immediately obvious upon visual inspection of typical field magnitude maps like the ones displayed in Fig.~\ref{fig:general}a-c. Scattering occurs at the chip package boundaries that act like a micro reverberation chamber as well as at interfaces of materials with different dielectric constants within the enclosure. The field maps display the speckle patterns characteristic of rich scattering and wave chaos~\cite{galdi2005wave}. The spatial extent of each speckle grain is $\Delta l_{\mathrm{corr}} \approx \lambda_0/2$. The wavelength $\lambda_0$ depends on the relative permittivity of the propagation medium which is nine times higher within the AlN layer than within the surrounding air. The speckle grains' approximately isotropic nature is direct evidence of the superposition of waves incident from all possible angles of arrival. Clearly, free-space intuition and concepts like beam-forming are not applicable in such a rich-scattering environment.

The field rapidly decorrelates not only as a function of spatial position (compare Fig.~\ref{fig:general}a and Fig.~\ref{fig:general}b) but also as a function of frequency $f_0$ (compare Fig.~\ref{fig:general}a and Fig.~\ref{fig:general}c). The spectral extent of a each speckle grain is $\Delta f_{\mathrm{corr}} \approx f_0/Q$, where $Q$ is the enclosure's composite quality factor. The latter essentially quantifies how often a ray bounces around the enclosure before becoming insignificant due to attenuation. $Q$ is hence directly related to the CIR duration. 
We evaluate $Q$ as $\pi f_0/\mu$, where $f_0= 60\ \mathrm{GHz}$ is the center of the considered frequency band (from 55~GHz to 65~GHz) and $\mu$ is the exponential decay constant of the average of the CIR magnitude envelopes. For a silicon layer of thickness $t=100\ \upmu$m that is considered in Fig.~\ref{fig:general}a-c, we obtained $Q=252$. As the thickness of the absorbing silicon layer is increased to 150~$\upmu$m and 200~$\upmu$m, the waves are attenuated quicker, the CIR becomes shorter, and we find $Q=74$ and $Q=56$, respectively. For even thicker silicon layers, the CIR is pulse-like without any significant tail~\footnote{The CIR's tail is known as ``coda'' or ``sona'' in wave physics.} such that no meaningful value of $Q$ can be determined.

\begin{figure}[b]
    \centering
    \includegraphics[width=\linewidth]{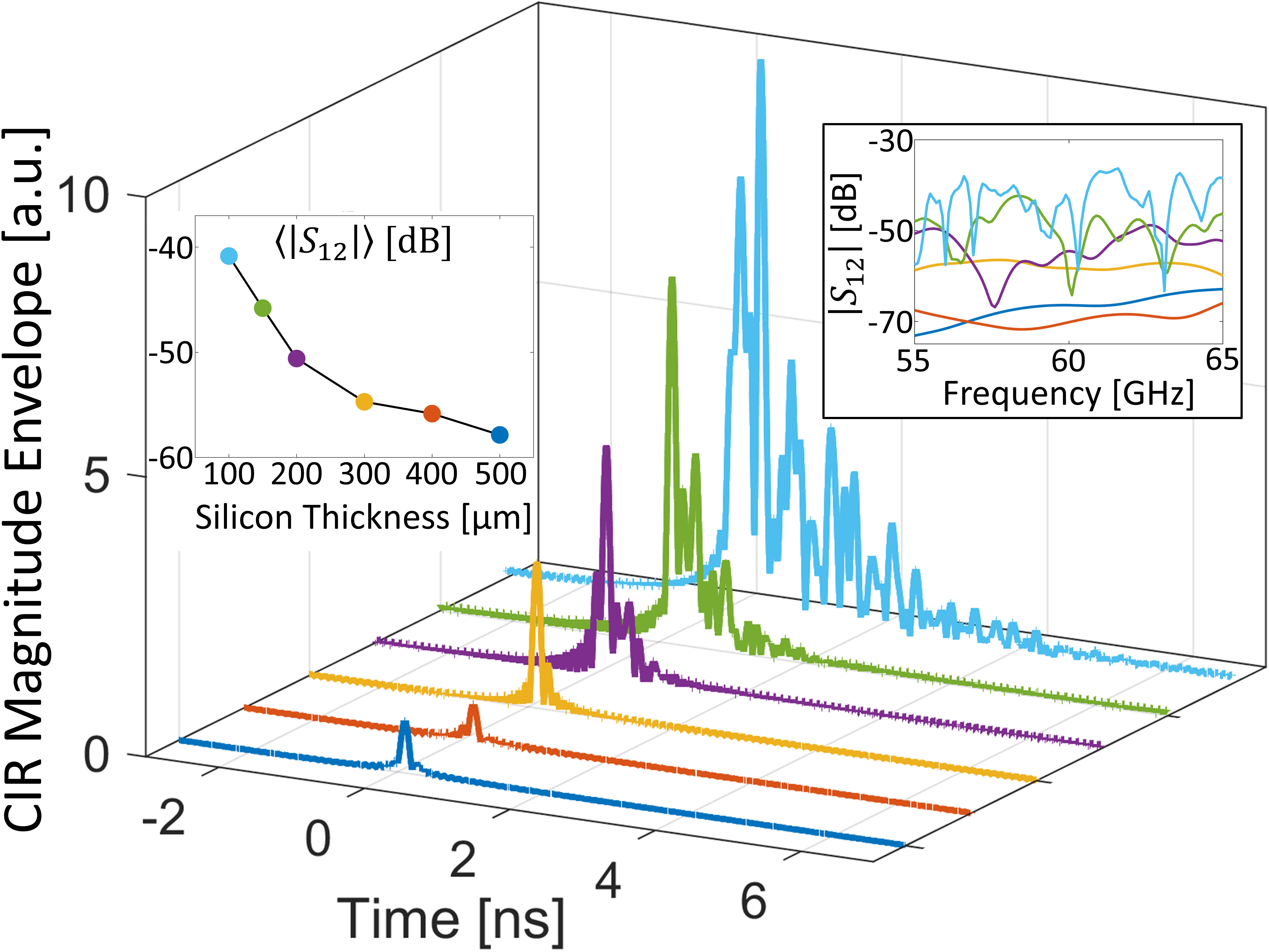}
    \caption{
    \textbf{Example CIR profiles for various silicon layer heights.} Different values of the silicon layer height are color-coded. The displayed CIRs correspond to the TX-RX3 transceiver pair shown in Fig.~\ref{fig:general}a (the two ports are separated by 9~mm). 
    The top right inset shows the corresponding channel spectra $S_{12}(f)$ over the considered $55\ \mathrm{GHz}<f<65\ \mathrm{GHz}$ interval. 
    The top left inset plots our pathloss metric $\langle|S_{12}|\rangle_f$ as a function of silicon layer thickness.
    }
    \label{fig2}
\end{figure}

At the same time, the longer the waves reverberate, the more energy accumulates inside the enclosure and the stronger is the received signal. A higher value of $Q$ corresponds both to more stored energy and a higher dwell time. In other words, a stronger RSSI is inevitably associated with a longer CIR, the latter directly implying more ISI in OOK modulation. We illustrate this RSSI-ISI dilemma in Fig.~\ref{fig2}, where we plot the CIRs between a transceiver pair separated by 9~mm for different values of the silicon layer's thickness~\footnote{\label{note1}The observation of non-zero signal already before $t<0$ in Fig.~\ref{fig2} is, of course, not violating causality but simply due to the finite bandwidth. The temporal position of the first peak corresponds to the time of flight of the shortest path. If the rise time is longer than the shortest path, inevitably the signal must begin to rise before $t=0$. Note also that the effect is enhanced in Fig.~\ref{fig2} because we used a rectangular rather than smooth window for the inverse Fourier transform.}. 
In addition, we show the corresponding channel spectra $S_{12}(f)$ and we quantify the pathloss as $\langle|S_{12}(f)|\rangle_f$. 

\textit{Remark 1:} The absolute values of this pathloss metric are relatively low because of the considered electrically small and near omni-directional slot antennas.

The on-chip RSSI-ISI dilemma is immediately obvious: for thick silicon layers ($t>250$~$\upmu$m), the spectrum is essentially flat and the CIR is almost pulse-like (similar to quasi-free space) but the RSSI is very low. For thinner silicon layers, the RSSI can increase by almost 20 dB but the spectrum displays more and more dispersion, resulting in lengthy CIRs. The latter evidence substantial reverberation and rich scattering inside the chip enclosure and pose a severe ISI challenge for OOK. The CIR maximum does not even have to coincide with the first CIR peak, as seen for the light-blue CIR in Fig.~\ref{fig2}. Indeed, the mean delay of the channel~\cite{goldsmith2005wireless} for the considered transmitter-receiver pair is 0.56~ns for the 100~µm thick silicon layer and drops down to 0.14~ns for silicon layers thicker than 250~µm. Hence, the majority of the signal energy arrives not along the shortest path but after significant reverberation at a later time for the scenarios with thin silicon layer. Note that in the absence of a line-of-sight, the shortest path cannot be related to the antenna separation.

\section{On-Chip RIS Design and Characterization}
\label{sec:design}

\begin{figure*}
    \centering
    \includegraphics[width=0.7\linewidth]{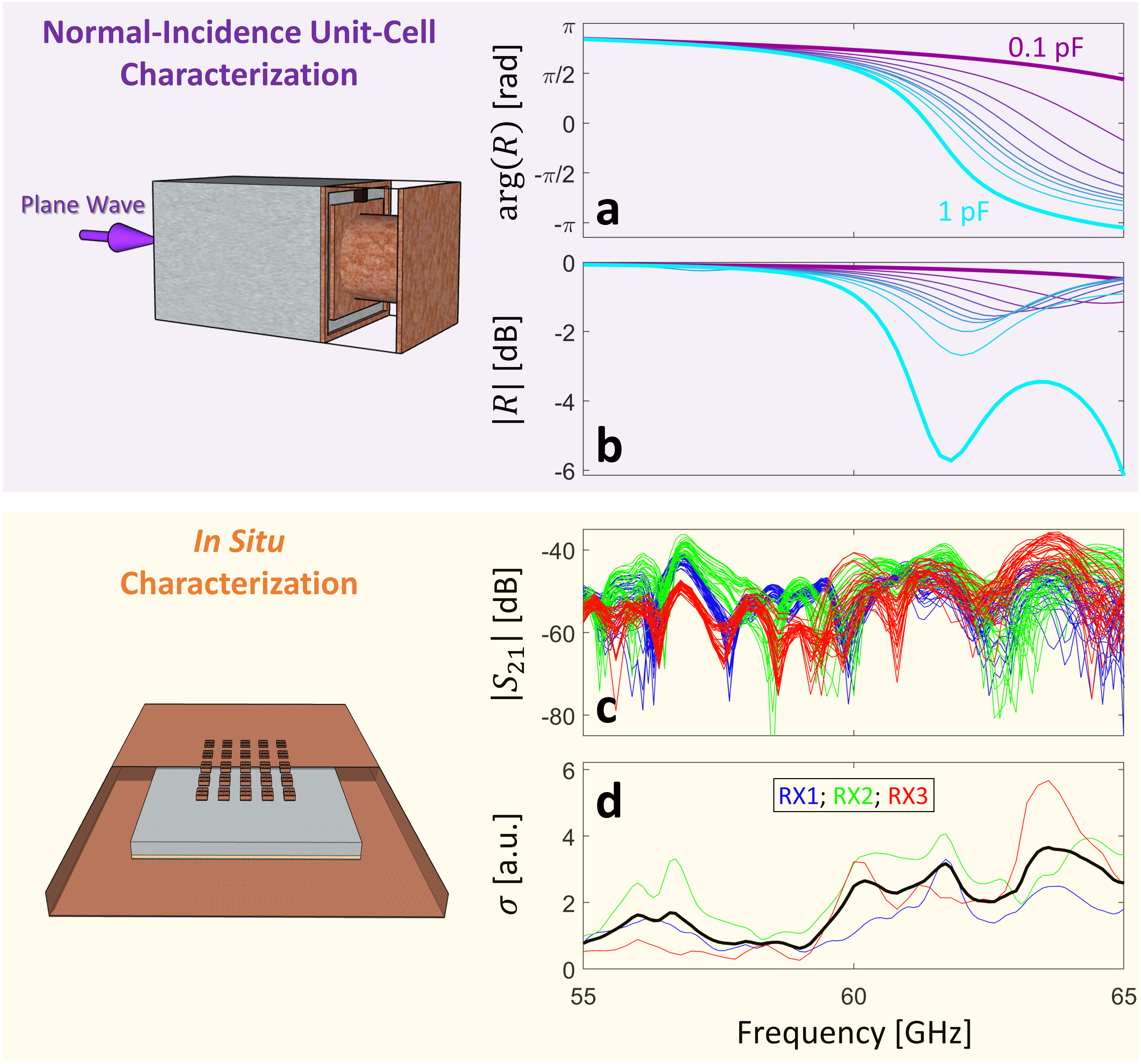}
    \caption{\textbf{Characterization of the on-chip RIS.} (a,b) Conventional normal-incidence unit-cell characterization method. A plane wave is normally incident on an infinite array of programmable meta-atoms, i.e., using periodic boundary conditions on all four sides of the meta-atom. Amplitude (a) and phase (b) of the reflection coefficient $R$ are shown for 10 values of the varactor capacitance.
    (c,d) \textit{In situ} characterization method. (c) Channel magnitudes $|S_{21}(f)|$ for three different receiver locations (color-coded, see Fig.~\ref{fig:general}a for locations) and 32 random RIS configurations evaluated directly in the considered WNoC setting. (d) Standard deviations $\sigma(f)$ of the complex-valued channels $S_{21}(f)$ for each pair of transceiver locations (color-coded), and averaged (black).}
    \label{fig3}
\end{figure*}

To eliminate the on-chip RSSI-ISI dilemma, we integrate a RIS into the ceiling of the chip package, as seen in Fig.~\ref{fig:general}d,e. The design of our programmable metasurface is detailed in Fig.~\ref{fig:general}f; the programmable meta-atom design is based on the well-known mushroom structure~\cite{Sievenpiper1999,sleasman2016microwave} equipped with a varactor diode which individually alters the effective capacitance of the meta-atom and shifts its resonance frequency. 
While the current RIS literature considers meter-scale applications that operate in air, our on-chip RIS must operate within an aluminium nitride layer whose high dielectric constant ($\sim\!8.8$) might require small varactors with very large effective capacitances. To avoid this problem, we design the meta-atom with one side to be air/vacuum (in future implementations a low-dielectric-constant substrate). Interestingly, such a configuration is the opposite of usual RIS designs where the metamaterial element is implemented on a high-dielectric-constant substrate and operates within air. 
Further details on our on-chip RIS design are presented in the Methods Sec.~\ref{subsec:RISdesign}. 

\textit{Remark 2:} Our generic proposal of shaping the CIRs of WNoCs with an on-chip RIS does not depend on the specific RIS design. Given the rich-scattering nature of the on-chip wireless environment, we do not seek to implement an analytically calculated surface impedance with high precision. Instead, the purpose of each RIS element is to maximally impact as many ray paths as possible, and using the iterative procedure detailed in Sec.~\ref{sec:shaping} and Sec.~\ref{subsec:optimization} we learn how to configure the RIS \textit{in situ}. Our algorithm adapts to any given properties of the RIS and the rich-scattering wireless environment.

We first characterize the fundamental building block of our on-chip RIS in the conventional manner: we study the reflection under normal-incidence illumination of an infinite array of our programmable meta-atom -- see Fig.~\ref{fig3}a,b. Magnitude and phase of the reflected wave display a resonant behavior in the targeted frequency range around 60~GHz; by changing the varactor's capacitance from 0.1~pF to 1~pF, this resonance can be tuned. 
However, this conventional unit-cell characterization method neglects the impact of coupling effects between neighboring RIS elements that can significantly alter the response. Indeed, as seen in Fig.~\ref{fig:general}d,e, we intend to deploy distributed $2 \times 2$ groups of synchronized meta-atoms, each group controlled by a single bias voltage line; we refer to these groups as meta-pixels. This $2 \times 2$ configuration will reduce the resonance frequency because the loading from the infinite array is removed. Overall, our on-chip RIS consists of a $5 \times 5$ array of such meta-pixels and covers 3.3~\% of the chip package ceiling, as seen in Fig.~\ref{fig:general}d,e. For simplicity we limit ourselves to 1-bit (binary) programmable meta-pixels in the following (0.1~pF or 1~pF).

To better understand the true potential for wave field manipulation with our on-chip RIS in our targeted on-chip application, we perform a second characterization \textit{in situ}. Specifically, we evaluate within the targeted on-chip environment shown in Fig.~\ref{fig:general} the channel $S_{12}$ between various pairs of antennas, each time for 32 random RIS configurations. The standard deviation $\sigma$ of these complex-valued channels across different configurations is a reliable metric of the ability of the RIS to manipulate these channels in the considered setting~\cite{RichScatteringRIS_magaz,Kaina2014SciRep,PhysFad}. The ability of our RIS to modulate frequencies across the entire considered $55-65\ \mathrm{GHz}$ band is now evident in Fig.~\ref{fig3}c,d.

We note that the presence of the on-chip RIS elements notably decreases the reverberation time of the on-chip enclosure because the on-chip RIS partially absorbs the waves that are impinging upon it. Indeed, we find that for a 100~$\upmu$m thick silicon layer the enclosure's quality factor drops from $Q=252$ to $Q=140$. 
Correspondingly, our pathloss metric $\langle | S_{12} | \rangle_f$ drops from $-41$~dB to $-51$~dB. Nonetheless, this remains the rich-scattering regime. 
As noted above, the physical mechanism of RIS-empowered over-the-air equalization relates to tailored constructive and destructive interferences rather than absorption. In fact, whether the additional absorption due to the presence of the on-chip RIS improves (shorter CIR) or deteriorates (lower RSSI) the wireless performance cannot be answered in general but depends on parameters such as the OOK modulation speed and the noise level, as discussed in Sec.~\ref{sec:communication}. The additional RIS-induced absorption could be reduced by designing an alternative RIS for which, for example, the resonances are moved away from the band of utilized WNoC frequencies.
The additional RIS-induced absorption can also be counter-balanced by adjusting the silicon layer's thickness. For instance, we find that our pathloss metric for a 100~$\upmu$m-thick silicon layer \textit{with} RIS is that of a 160~$\upmu$m-thick silicon layer \textit{without} RIS.

\section{On-Chip CIR Shaping}
\label{sec:shaping}

\begin{figure}[b]
    \centering
    \includegraphics[width=0.9\linewidth]{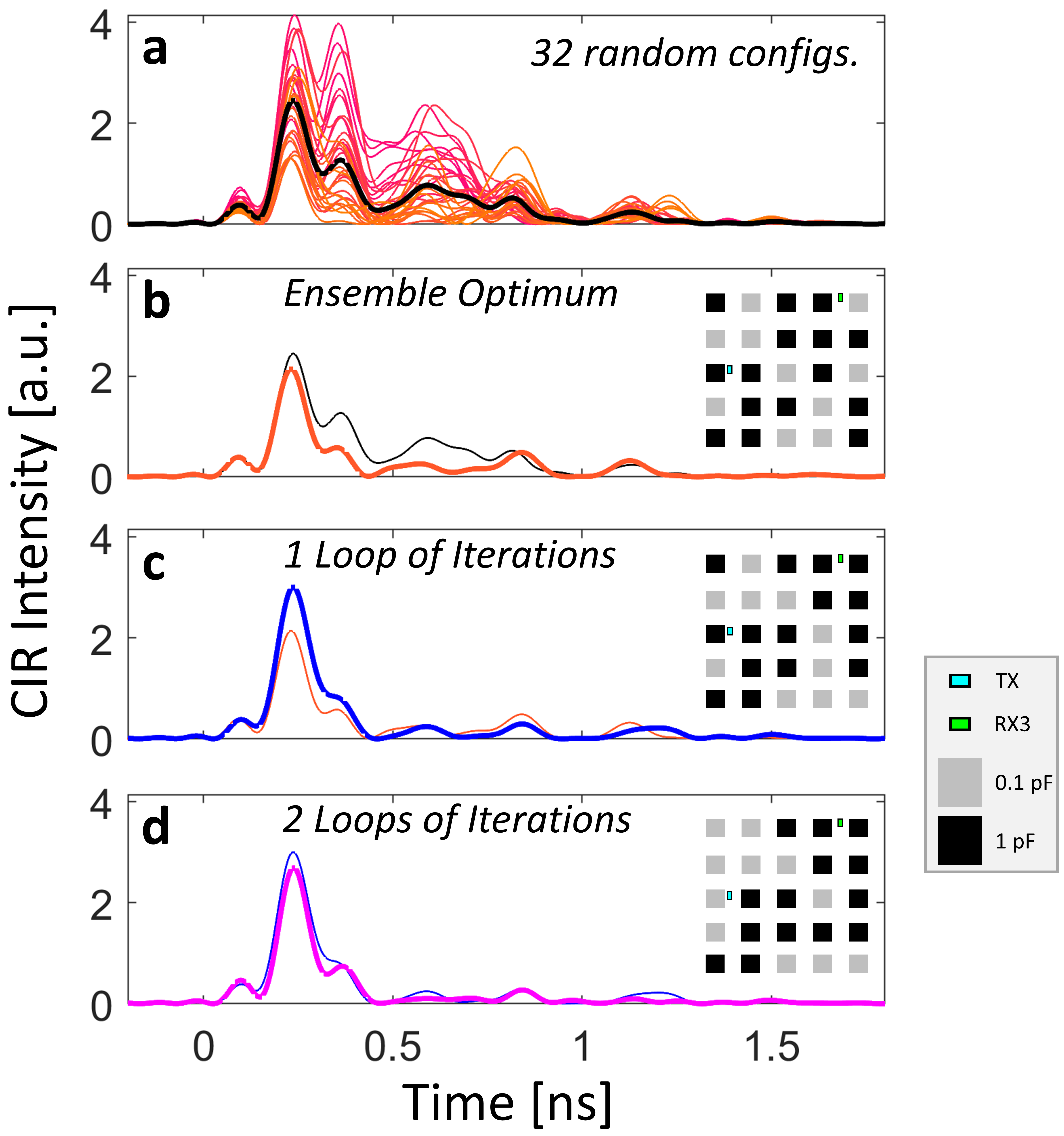}
    \caption{\textbf{CIR optimization with on-chip RIS.} (a) CIR intensity profiles for 32 random on-chip RIS configurations (color-coded), and the average CIR (black). (b) Best out of the 32 random configurations. (c) Optimum after one optimization loop. (d) Optimum after two optimization loops. The insets indicate the corresponding RIS configuration and antenna locations.}
    \label{fig4}
\end{figure}

We now proceed with optimizing the on-chip RIS configuration to achieve the desired CIR shape, namely a pulse-like CIR despite rich scattering. As detailed in the Methods Sec.~\ref{subsec:optimization}, first, we define a cost function $\mathcal{C}$ that quantifies the portion of energy in the strongest CIR tap relative to the CIR's total energy; second, we use an iterative optimization to identify a RIS configuration that optimizes $\mathcal{C}$. To expedite the search within the huge search space ($2^{25}$ possible RIS configurations), we determine a reasonably good initial guess: the best out of 32 random RIS configurations. Then, we flip the state of one meta-pixel at a time, and keep the change of its state if the CIR has become more pulse-like according to our criterion. 
It is not sufficient to test each meta-pixel only once because the presence of reverberation inside the chip enclosure correlates the optimal states of different meta-pixels~\cite{del2016spatiotemporal} -- see Fig.~\ref{fig:OptiDynamics}b.

\begin{figure}[b]
    \centering
    \includegraphics[width=\linewidth]{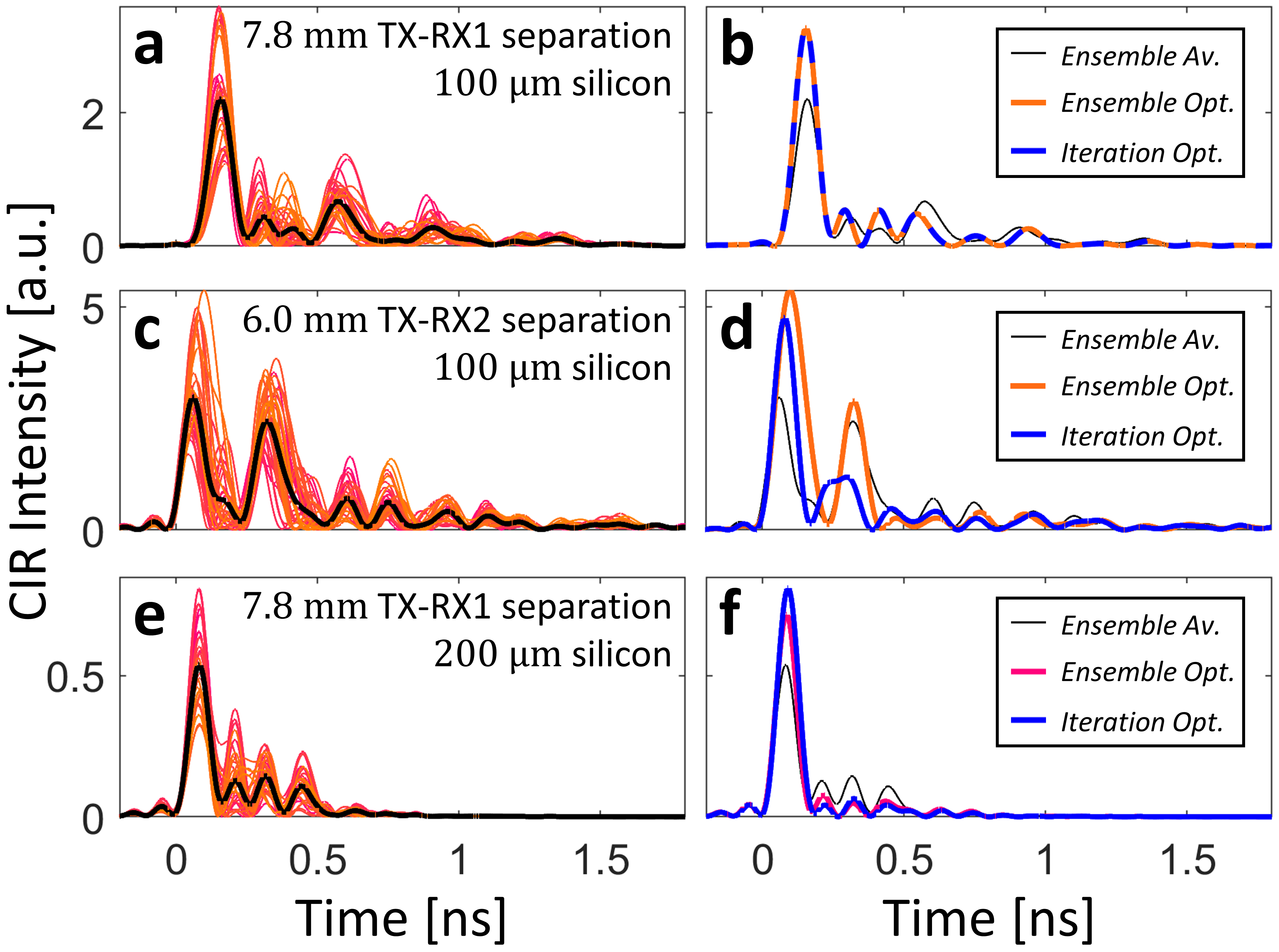}
    \caption{\textbf{Further examples of CIR shaping.} Different transceiver separations and silicon layer thicknesses are considered -- see insets in (a,c,e). CIR intensity profiles for 32 random realizations (a,c,e) as well as the ensemble average (black), ensemble optimum (red) and iteration optimum (blue) are shown (b,d,f).}
    \label{fig5}
\end{figure}

An example outcome of this protocol for on-chip CIR shaping is displayed in Fig.~\ref{fig4} for the case of a 100~$\upmu$m thick silicon layer and 9~mm distance between transmitter and receiver (TX-RX3). Our optimization progressively flattens the CIR behind the first peak, such that the modulation rate for OOK-based data transmission between the ports can be increased substantially. It is also apparent that for the optimized configurations roughly half of the RIS elements are in one and the other half in the other one of the two possible states, without any intuitively understandable pattern. This is expected in a rich-scattering system and in contrast to the (analytically calculated) RIS patterns used for beamforming in free space~\cite{cui2014coding}.

To illustrate the generality of our technique, as well as its real-time \textit{in situ} adaptability to dynamic traffic patterns~\cite{jog2021one}, we apply it also to two other pairs of antenna locations on a chip with 100~$\upmu$m thick silicon layer as well as to one case on a chip with 200~$\upmu$m thick silicon layer. The effectiveness of our technique is clearly seen for all three cases in Fig.~\ref{fig5}. In the first case, the iterative optimisation does not yield any improvement upon the ensemble optimum. In the third case with 200~$\upmu$m thick silicon layer, the CIR is, of course, shorter but our technique is effective nonetheless in improving the CIR profile toward a pulse-like shape for OOK purposes. Overall, these three further examples confirm that depending on traffic needs on the chip, the CIR of a selected antenna pair can be shaped to be pulse-like, irrespective of antenna separation and silicon layer thickness.

Before closing this section, we underline the relevance of ergodicity in wave chaos for the generality of our results. According to the concept of ergodicity, different realizations of the considered system are governed by the same statistics. Therefore, in a system that is merely a different realization of our prototypical system (geometrically modified while keeping global parameters like $Q$ constant), one can faithfully expect to achieve similar amounts of wave control with the same RIS (of course, the optimal RIS configurations would be a different one). Thus, simplifications in our prototypical model do not impact the generality of our conclusions.

\section{Communication Analysis}
\label{sec:communication}

Having sculpted the CIRs into the desired pulse-like shape despite severe multipath in Sec.~\ref{sec:shaping}, we now proceed to evidencing the ensuing advantages in the targeted OOK communication scheme in terms of the BER. To be an enticing alternative to wired interconnects, WNoCs would have to reach BERs below $10^{-15}$ at symbol rates of at least 10~Gb/s~\cite{timoneda2020engineer}. A rigorous evaluation of the BER consists in simulating the transfer of many symbols at a specific noise level and modulation speed in order to determine how many symbols were received incorrectly. It is clear that such a rigorous analysis cannot be conducted for $10^{15}$ or more symbols due to the associated computational cost, and therefore it is impossible to confirm BERs below $10^{-15}$ through rigorous analysis.

A commonly used work-around is to assume that simple analytical expressions for the BER are applicable~\cite{timoneda2020engineer}, including a treatment of ISI as random noise. However, ``ISI noise'' exhibits strong correlations; moreover, because for BERs below $10^{-15}$ we would be interested in extreme outliers of a statistical distribution (the one incorrectly transmitted symbol out of $10^{15}$ symbols), it is impossible to confirm that such approximations accurately describe the physical reality by looking at statistical moments like the mean or the variance. Therefore, instead of seeking an \textit{approximative quantitative} analysis that confirms BERs below $10^{-15}$, here we opt for a  \textit{rigorous qualitative} BER analysis to evidence the advantages of our sculpted CIRs. 
We focus on regimes with higher BERs (around $10^{-2}$) where we can robustly estimate the BER with our rigorous analysis based on ``only'' $2.28\times 10^5$ transmitted symbols (see Methods Sec.~\ref{subsec:ook}). Consequently, the reader's attention should be on the qualitative changes of the BER curves that we present rather than on specific BER values. In any case, the quantitative BER results could be improved (i) through techniques such as return-to-zero modulation and threshold adaptation~\cite{timoneda2020engineer}, and (ii) by using carefully designed antennas (instead of electrically small ports). If well-matched antennas were used, the amount of radiated and captured energy would increase; thereby the RSSI and ultimately the SNR would increase for any given combination of modulation speed and noise level, thus improving the BER.

\begin{figure*}
    \centering
    \includegraphics[width=\linewidth]{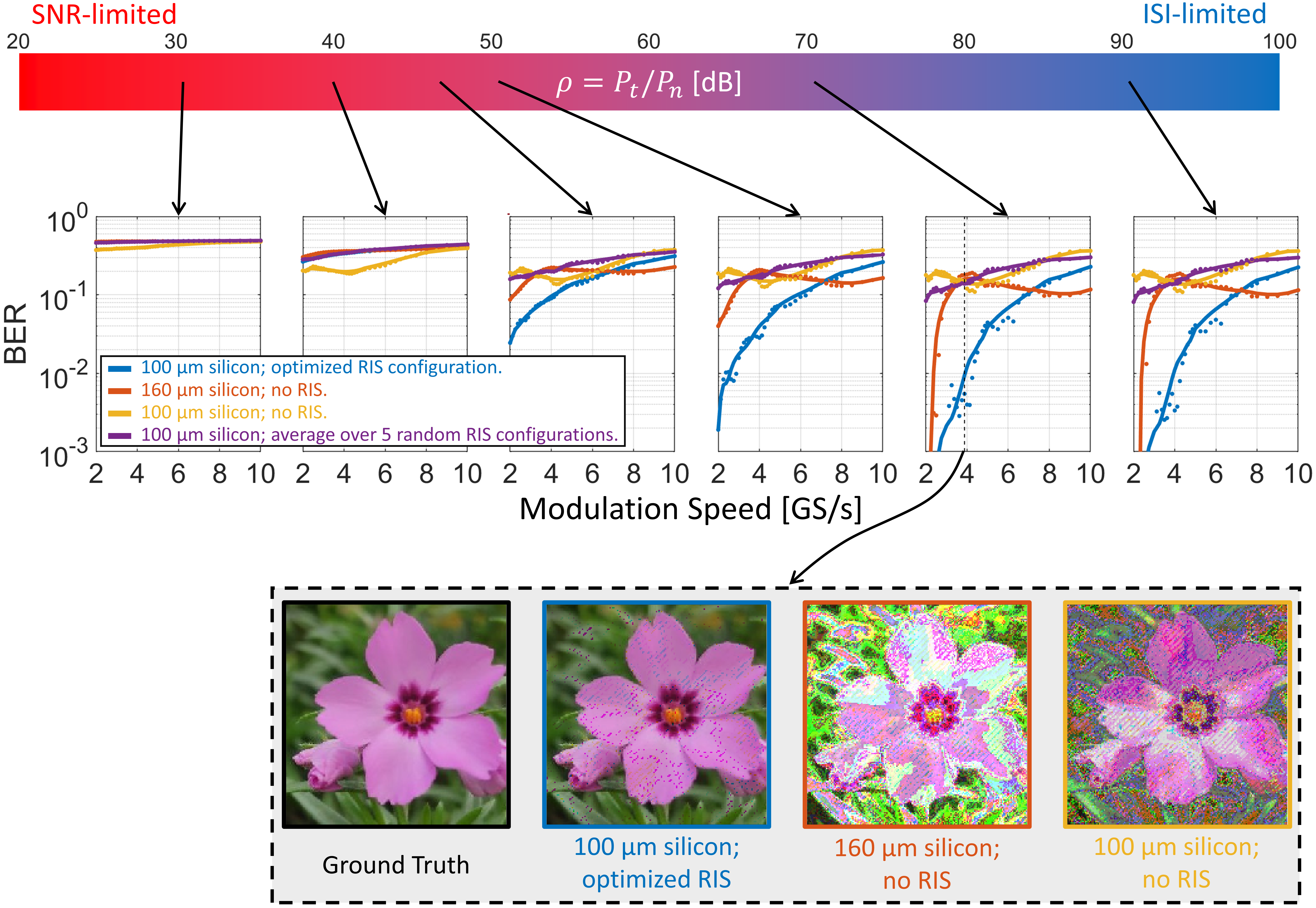}
    \caption{\textbf{BER in OOK on-chip communication.} The top bar indicates the considered range of the ratio $\rho=P_t/P_n$ between transmitted power $P_t$ and noise power $P_n$. Note that $\rho$ is \textit{not} the SNR at the receiver because the latter depends on the pathloss which differs in the considered scenarios. For six representative values of $\rho$, we plot the dependence of the BER on the modulation speed for the TX-RX3 wireless on-chip link. We consider four scenarios: (blue) 100~$\upmu$m silicon layer \textit{with optimized} RIS configuration; (red) 160~$\upmu$m silicon layer \textit{without} RIS (same pathloss as blue); (yellow) 100~$\upmu$m silicon layer \textit{without} RIS; (purple) 100~$\upmu$m silicon layer \textit{with random} RIS configuration, averaged over five random RIS configurations. The raw BER results based on $2.28\times 10^5$ symbols are indicated with dots, while the continuous line is a smoothed version thereof. We further illustrate the benefits of RIS-empowered CIR shaping for $\rho=70$~dB and a modulation speed of 3.875~GS/s. We simulate the transfer of a photographic image of a \textit{phlox subulata} flower taken in Alzenau, Germany, for the first three considered scenarios. The ground truth and the three received images are shown.
    }
    \label{fig:comm}
\end{figure*}

For concreteness, we focus on the sculpted CIR from Fig.~\ref{fig4} for our OOK communication analysis in Fig.~\ref{fig:comm}~\footnote{While the curves in Fig.~\ref{fig:comm} clearly show the general trend, they are to some extent specific to a single wireless on-chip link (TX-RX3). This lack of complete universality explains why they are not perfectly smooth and monotonous: on a few occasions, slightly increasing the modulation speed may actually slightly improve the BER due to the specific CIR structure. Upon averaging such curves over different wireless on-chip links, such non-universal features would be averaged out.}; recall that in this setting the chip's silicon layer is 100~$\upmu$m thick. To meaningfully interpret the dependence of the sculpted CIR's BER in OOK-based communication on modulation speed and noise level, we identify the following three benchmarks:

\begin{itemize}
  \item A chip \textit{without} RIS and 160~$\upmu$m thick silicon layer, because the pathloss for the considered channel, quantified as $\langle |S_{21}(f)| \rangle_f$, is the same as for our chip \textit{with} RIS and 100~$\upmu$m thick silicon layer. The additional silicon thickness thus mimics the additional absorption originating from the presence of a RIS.
  \item A chip \textit{without} RIS and 100~$\upmu$m thick silicon layer, that is, the chip we consider without adding the RIS. The pathloss is thus less because the additional absorption due to the RIS is missing.
  \item A chip \textit{with randomly configured} RIS and 100~$\upmu$m thick silicon layer. Specifically, we report the average BER based on five random RIS configurations. This benchmark helps to determine if potential BER improvements are due to the mere presence of the RIS (this benchmark) or rather its judicious configuration (the result from Fig.~\ref{fig4}).
\end{itemize}

Besides the CIR, the BER also pivotally depends on the OOK modulation speed and the noise level. Because the pathloss depends on the wireless channel, we quantify the noise level in terms of the ratio of transmitted power $P_t$ to noise power $P_n$, namely $\rho=P_t/P_n$, as opposed to using the SNR of the received signal. As seen in Fig.~\ref{fig:comm}, the noise level defines two extreme regimes. For low values of $\rho$, the communication is SNR-limited: no matter what wireless channel and modulation speed is used, the BER almost reaches its upper bound of 0.5. At the other extreme, for high values of $\rho$, OOK communication is ISI-limited: in this regime, wireless communication can be successfully operated with low BER if wireless channel and modulation speed are chosen appropriately. In this regime, for a given wireless channel, lowering the modulation speed reduces the BER because it mitigates ISI. Moreover, differences between the considered wireless channels are very significant in this regime.

As soon as $\rho$ is sufficiently large for the communication not to be SNR-limited, the benefits of CIR shaping manifest themselves. For the examples in Fig.~\ref{fig:comm} with medium or high $\rho$, the blue curve corresponding to the RIS-optimized CIR lies significantly below the three benchmarks. For concreteness, consider that our targeted maximum BER is $10^{-2}$ for $\rho=70\ \mathrm{dB}$. Then, the optimized RIS configuration would allow us to operate at 4~GS/s, which is twice the maximum modulation speed at which the best out of the three benchmarks (red) could guarantee a maximal BER of $10^{-2}$. We visualize these differences by emulating the transfer of a color image across the considered channels at $3.875~$GS/s and $\rho = 70$~dB. While the RIS-shaped channel transmits the image almost flawlessly, the benchmark channels yield heavily distorted received images. A successful transmission of the image via the benchmark channels would require a much lower modulation speed.

From these results we conclude that

\begin{enumerate}[label=(\roman*)]
  \item it is the judicious configuration of the RIS as opposed to its mere presence in a random configuration that yields the substantial performance improvements (blue vs purple curves). 

\item absorption by the RIS is not a significant effect because benchmarks with the same pathloss perform much worse (blue vs red curves).

\item lower pathloss is by no means a guarantee for a lower BER except for when the wireless channel is SNR-limited. Indeed, the wireless channel with the lowest pathloss (yellow) performs best (but still poorly) only for low values of $\rho$.

\end{enumerate}

\noindent All three points underline that the  mechanism underlying our CIR shaping is not related to the additional absorption introduced by the presence of the RIS in the on-chip environment. Instead, as outlined in the Introduction, by controlling the delays of different paths, we create constructive (destructive) interferences at (before/behind) the main CIR tap.

\section{Discussion}
\label{sec:discussion}

The ability of the RIS to shape the CIR depends on several factors and can hence be improved accordingly. The most important question is, how many of the rays that connect the transmitter to the receiver have encountered the RIS? The percentage of rays manipulated by the RIS can be increased by (i) using more RIS elements (currently our RIS covers only 1.5~\% of the interior chip package surface), (ii) placing the RIS elements closer to the transceivers, and/or (iii) operating in an environment with more reverberation. Moreover, the design of the RIS elements can be optimized further, and linear instead of binary control would also enhance the available control over the EM field. In addition, linear programmability would help to deploy more efficient gradient-descent optimization algorithms to search the huge optimization space. A further consideration regards the EM field polarization. Assuming that the field inside the chip enclosure is approximately chaotic, which is a reasonable assumption when the silicon layer is thin, means that all polarizations are well mixed and the restriction of our meta-atom to engage mainly with one polarization is not problematic, as evidenced by our results; nonetheless, by alternating the varactors on the meta-atoms constituting the meta-pixels as in Ref.~\cite{sleasman2016microwave}, or using multiple varactors on each of them, modifications of our on-chip RIS design that engage with both field polarizations can be conceived.

In contrast to the meter-scaled smart indoor environments that dynamically evolve due to the motion of inhabitants, smart on-chip EM environments are sealed and extraordinarily static such that suitable RIS configurations for various traffic patterns can be identified in a one-off calibration phase. However, the influence of temperature fluctuations on the wireless channels remains to be investigated; the dielectric constant of the different chip layers may vary with temperature such that the wireless on-chip channels could evolve. A significant temperature dependence would call for a self-adaptive on-chip RIS that updates its configuration based on a current temperature estimate~\cite{ChloeHelsinki}. Such an estimate could be derived from knowledge of the current chip activity or based on a simple temperature sensor.

\section{Conclusions and Outlook}
\label{sec:conclusion}

To summarize, we demonstrated that the integration of a RIS into an on-chip wireless environment can endow the latter with programmability, a functionality that we leveraged to achieve pulse-like CIRs despite rich scattering. Thereby, we overcame the RSSI-ISI dilemma that plagues current WNoC proposals: we mitigated ISI without simultaneously reducing the RSSI. Pulse-like CIRs enable faster data transmission rates in OOK communication because they equalize the wireless on-chip channel ``over-the-air''. 
Our rigorous OOK communication analysis confirmed that RIS-shaped CIRs can double the permissible modulation speed for a given desired BER value at a given noise level.

Looking forward, more complex communication scenarios such as communication from one transmitting to multiple receiving nodes (SIMO) deserve attention~\cite{del2019optimally,imani2021demand,phang2018near}. Ultimately, experimental validation of all these ideas will be indispensable in the future. 
Moreover, our proposed smart on-chip EM environment can be endowed with a second functionality related to wave-based analog signal-processing by bringing recent proposals of wave processing in (programmable) scattering enclosures for matrix multiplication~\cite{del2018leveraging}, signal differentiation~\cite{sol2021meta}, or reservoir computing~\cite{ma2021short} to the chip scale. Such analog ``over-the-air'' computing holds the promise to be faster and more energy efficient than its electronic digital counterpart for specific computational operations, paving the way to hybrid analog-digital processing chips. We also foresee the possibility of communication-efficient RIS-empowered on-chip federated learning for the collaboration of different cores on the same chip to train a machine-learning model~\cite{liu2021reconfigurable,ni2021federated}.

A further avenue for future exploration is to consider similar problems of data exchange inside rich-scattering enclosures at intermediate scales between the chip scale and the indoor scale. Relevant examples include communication inside racks or blades, inside the chassis of personal computers, or inside data centers~\cite{chiang2010short,zhan2019low,fu2020modeling}.

\section{Methods}
\label{sec:methods}

\subsection{Full-Wave Simulation Setup}
\label{subsec:cst}

We accurately model and characterize the on-chip environment with Ansoft's High Frequency Structural Simulator (HFSS), a field solver based on the finite-element method. HFSS simulates wave propagation in the three-dimensional on-chip environment and produces the scattering parameters associated with the connected ports. We use HFSS' automatic mesh generator. HFSS outputs the channel $S_{12}(f)$ between the two ports in the frequency domain, and we use a Fourier transform to obtain the corresponding CIR~[73].

The solder bumps on the bottom and the chip package on the top and sides are taken to be copper which is modelled as (lossy) conductive surface in HFSS. Because the entire simulation domain is surrounded by this copper layer, no further boundary conditions remain to be specified. 
Given that the granularity of the solder bump layer is deeply sub-wavelength, it is common to describe this as solid metallic layer~\cite{timoneda2020engineer}. 
The high complexity of the wireless on-chip environment that gives rise to speckle-like wave fields (see Fig.~\ref{fig:general}a-c) implies that we are operating in the rich-scattering regime in which the concept of ergodicity guarantees that  different realizations will be governed by the same statistics. Specifically, this means that in a wireless on-chip environment that differs from ours in some geometrical detail, a similar degree of channel shaping can be achieved, while, of course, the optimal RIS configuration will be a different one. Therefore, the very complex deeply sub-wavelength metallic
patterning of the solder bumps does \textit{not} affect any of our conclusions.

The thickness and relative permittivity of the chip layers are summarized in Table~1. Following the standard HFSS library, silicon is taken to have a conductivity of $10\ \mathrm{S}\cdot \mathrm{m}^{-1}$ while silicon dioxide and aluminium nitride are taken to be lossless dielectrics. 

\begin{table}[h]
\caption{\label{table_singleDoF} Thickness and relative permittivity of the chip layers (see Fig.~\ref{fig:general}e).}
\begin{ruledtabular}
\begin{tabular}{cccccc}
Material &  Thickness  & Relative   \\
&[mm] & Permittivity\\
\hline
Silicon dioxide ($\text{SiO}_2$) & 0.011 & 4 \\
Silicon (Si) & $t$ & 11.9 \\
Aluminium nitride (AlN) & 0.8 & 8.8 \\

\end{tabular}
\end{ruledtabular}
\end{table}

The central wavelength at 60~GHz is 5~mm in air and 1.7~mm in AlN. The ports are implemented as slot antennas through $200\ \upmu\mathrm{m} \times 300\ \upmu\mathrm{m}$ apertures in the bottom metallic layer. Within the considered frequency band from 55~GHz to 65~GHz, the ports are thus electrically small and their radiation properties are essentially frequency-independent.

\subsection{On-Chip RIS Design}
\label{subsec:RISdesign}

Our designed on-chip RIS consists of a $5 \times 5$ array of meta-pixels, each consisting of a $2 \times 2$ group of programmable meta-atoms operating around 60~GHz. Our on-chip RIS covers 3.3~\% of the chip package ceiling and roughly 1.5~\% of the total interior surface of the metallic chip package. 

The design of the programmable meta-atoms is shown in Fig.~\ref{fig:general}f and based on typical mushroom structures loaded with varactor diodes. For simplicity, we limit ourselves to binary (1-bit) programmability. The two effective capacitances of the varactor diodes (0.1~pF and 1~pF) are selected to be within the range of typical GaAs junction diodes currently available off-the-shelf that can operate up to at least 70~GHz. In our full-wave simulations, we also include losses for the varactor diodes, assuming they have a quality factor of around 1000. In future experimental implementations, the programmable component of the meta-atom may be integrated or fabricated onto the chip (instead of using packaged off-the-shelf components). Furthermore, the choice of programmable component is not limited to varactor diodes. PIN diodes or Schottky diodes as well as field-effect transistors (FETs)~\cite{venkatesh2020high}, including the case of thin-film transistors, can also be used. Another option that works well at mmW and higher frequencies are meta-atoms equipped with liquid crystals. A further possibility may consist in leveraging piezoelectric diaphragms to mechanically alter the chip package boundaries~\cite{dirdal2022mems}.

\begin{figure}[h]
    \centering
    \includegraphics[width=\linewidth]{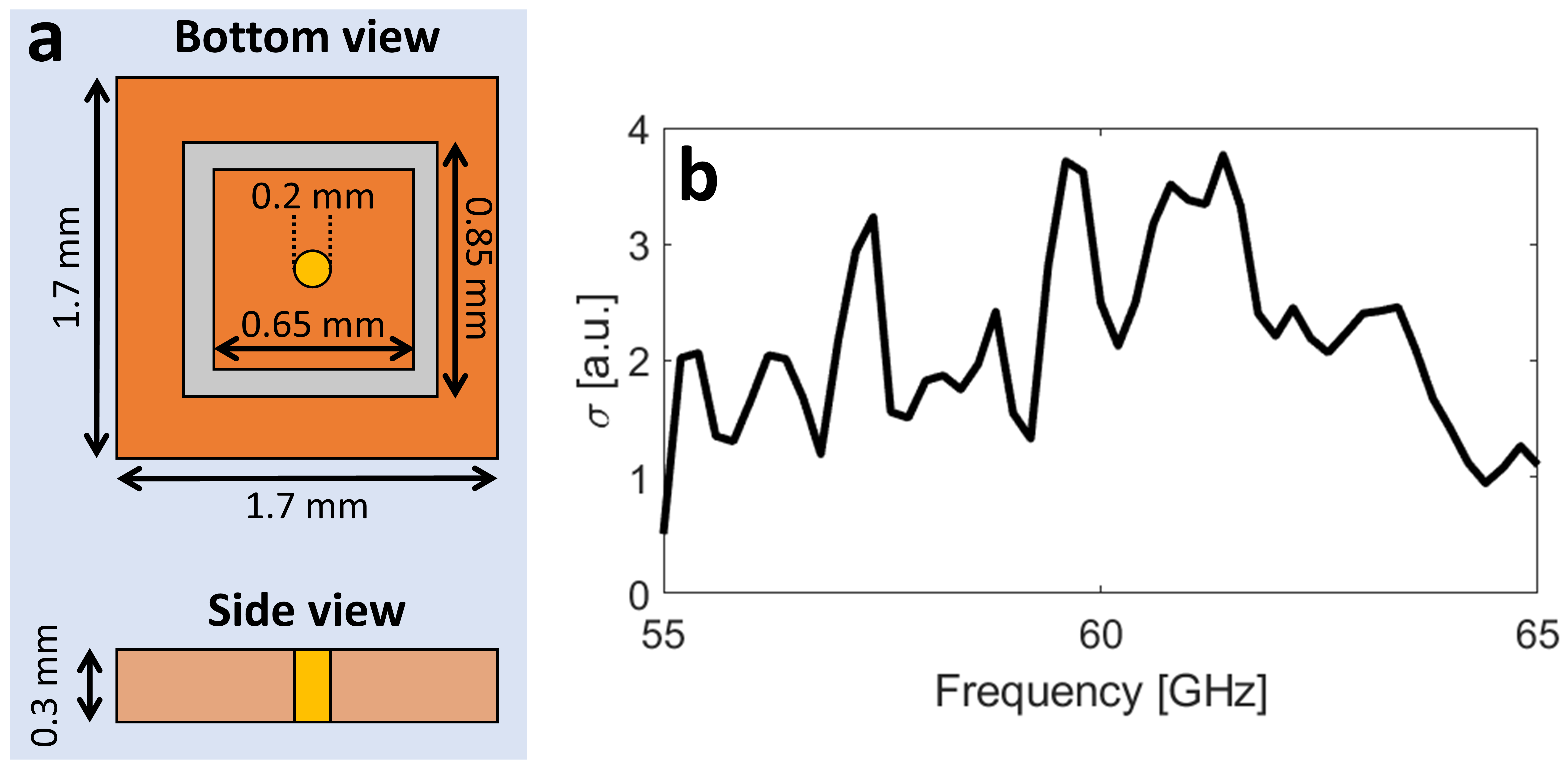}
    \caption{Alternative on-chip RIS design with an off-the-shelf packaged varactor diode. (a) Design outline. (b) \textit{In situ} characterization akin to Fig.~\ref{fig3}d.}
    \label{fig:AltDesign}
\end{figure}

If off-the-shelf components are to be used in the meta-atom design, it is critical that the latter allows one to incorporate the packaged varactor diodes. To that end, the meta-atom size would have to increase relative to that presented in Fig.~\ref{fig:general}f. To demonstrate that the use of packaged off-the-shelf varactors is possible, we have have designed an alternative programmable meta-atom that can include an off-the-shelf Macom varactor diode (MAVR-000120-1411). A schematic drawing of this design is presented in Fig.~\ref{fig:AltDesign}a and shows that its size exceeds that of the meta-pixels from Fig.~\ref{fig:general}. To confirm that this alternative on-chip RIS design (compatible with an off-the-shelf varactor diode) is also efficient at parametrizing the on-chip wireless channels, we replaced each meta-pixel in Fig.~\ref{fig:general} with one of the alternative meta-atoms in order to repeat the \textit{in situ} characterization from Fig.~\ref{fig3}c,d. The result in Fig.~\ref{fig:AltDesign}b shows clear variations of the wireless channel for random configurations of the alternative on-chip RIS.

Finally, in moving toward experimental realizations, the biasing network would have to be included. Biasing lines are not expected to severely impact the performance, and refined meta-atom designs taking the biasing lines into account can be conceived. As explained in Sec.~\ref{sec:design}, our generic proposal for shaping CIRs does not depend on the specific RIS design.

\subsection{Determination of RIS Configuration}
\label{subsec:optimization}

We define our cost function $\mathcal{C}$ that is to be maximized as the ratio of signal intensity in the strongest tap of the CIR to the total signal intensity:
\begin{equation}
\mathcal{C} = \frac{\int_{t_0-\Delta t/2}^{t_0+\Delta t/2} h^2(t) \mathrm{d}t}{\int_0^\infty h^2(t) \mathrm{d}t},
\label{eq:CF}
\end{equation}
where $t_0$ is the peak time and $\Delta t$ is the width of the strongest CIR tap.

Identifying a RIS configuration that maximizes $\mathcal{C}$ is a non-trivial task because (i) due to the rich-scattering geometry of the chip enclosure no analytical forward model exists that maps a given $\mathcal{C}$ to the corresponding CIR $h(t)$, and (ii) we limit ourselves to binary RIS configurations which are not compatible with standard gradient-descent optimization. Surrogate forward models can be learned if enough labeled training data is available~\cite{ChloeHelsinki}. Here, we opt for the alternative route of using the simple iterative Algorithm~1 similar to those used, for example, in Refs.~\cite{del2016spatiotemporal,PhysFad}. 

\begin{algorithm}[h]
	\caption{Binary RIS Optimization for On-Chip Over-the-Air Equalization}
	Evaluate $\{\mathcal{C}_i\}$ for $32$ random RIS configurations $\{C_{i}^0\}$.\\
	Select configuration $C_{\rm curr}$ corresponding to $\mathcal{C}_{\rm curr} = \mathrm{min}_i(\{\mathcal{C}_i\})$.\\
	\For{$i=1,2,\ldots,2N_{\rm RIS}$}{
	    Define $C_{\rm temp}$ as $C_{\rm curr}$ but with configuration of $\mathrm{mod}(i,N_{\rm RIS})$th RIS element flipped.\\
	    Evaluate $\mathcal{C}_{\rm temp}$.\\
	    	        \If{$\mathcal{C}_{\rm temp} > \mathcal{C}_{\rm curr}$}{
	            Redefine $C_{\rm curr}$ as $C_{\rm temp}$ and $\mathcal{C}_{\rm curr}$ as $\mathcal{C}_{\rm temp}$.
	        }
	}
	\KwOut{Optimized RIS configuration $C_{\rm curr}$.}
	\label{alg:opti}
\end{algorithm}

\begin{figure}[h]
    \centering
    \includegraphics[width=\linewidth]{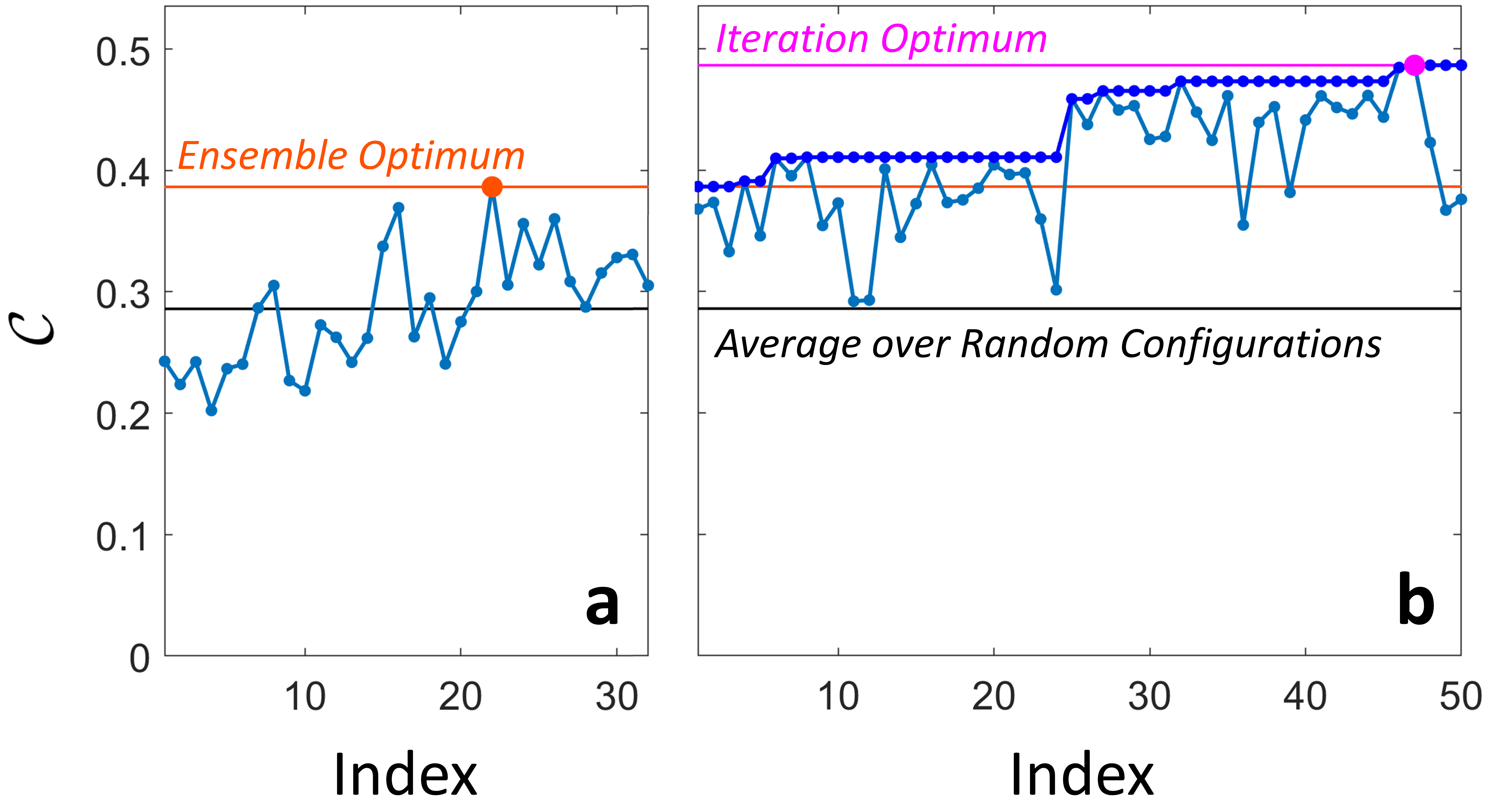}
    \caption{Optimization dynamics of Algorithm~1 for the case shown in Fig.~\ref{fig4}. (a) Step 1 of Algorithm~1: Evaluation of 32 random RIS configurations. (b) Step 2 of Algorithm~2: Iterative optimization of the RIS elements, looping twice over each element.}
    \label{fig:OptiDynamics}
\end{figure}

In a first step, we evaluate $\mathcal{C}$ for 32 random RIS configurations and identify the best one. The latter then serves as starting point for an iterative search. Therein, at every iteration the configuration of one RIS element is flipped to check if $\mathcal{C}$ increases, in which case the change is kept. We loop two times over each pixel because reverberation induces long-range correlations between the optimal configurations of individual RIS elements. Each evaluation of $\mathcal{C}$ for a new RIS configuration requires a new full-wave simulation to obtain the corresponding CIR. The detailed optimization dynamics of Algorithm~1 are illustrated for the case of Fig.~\ref{fig4} in Fig.~\ref{fig:OptiDynamics}.

\subsection{OOK Modulation Scheme}
\label{subsec:ook}

In this section, we detail our procedure to simulate an OOK modulation scheme. A high-level overview is provided in Algorithm~2. 
Based on the frequency-dependent channel $S_{21}(f)$ obtained from the full-wave simulations, we leverage the linearity of the wave equation to faithfully emulate the transfer of information with OOK, that is, binary amplitude-shift keying (BASK). 
The stream of $N$ bits that we intend to transfer is a $1\times N$ vector $\mathbf{b_t}$. We denote the $i$th entry of $\mathbf{b_t}$ by $\mathbf{b_t}[i]$ in the following.

\begin{algorithm}[h]
	\caption{OOK Data Transfer Simulation (\textit{high level overview})}
	Identify transmitted data stream $\mathbf{b_t}$.\\
	Determine transmitted signal $s_{\mathrm{TX}}(t)$ using Eq.~\eqref{eq:s_TX}.\\
	Determine received signal $s_{\mathrm{RX}}(t)$ using Eq.~\eqref{eq:s_RX}.\\
	Determine received data stream $\mathbf{b_r}$ using Eq.~\eqref{eq:d} and Eq.~\eqref{eq:b_r}.
	\label{alg:OOK}
\end{algorithm}

First, we create the transmitted signal $s_{\mathrm{TX}}(t)$ by modulating the amplitude of a sinusoidal carrier at $f_0 = 60~\mathrm{GHz}$ with the data stream $\mathbf{b_t}$ for a given choice of bit duration $\tau$:

\begin{equation}
s_{\mathrm{TX}}(t)\bigg\rvert_{
(i-1)\tau \leq t < i \tau} = \begin{cases}
0 &\text{if $\mathbf{b_t}[i]=0$.}\\
a \ \mathrm{sin}(2\pi f_0 t) &\text{if $\mathbf{b_t}[i]=1$.}
\end{cases}
\label{eq:s_TX}
\end{equation}

\noindent The modulation speed is hence $1/\tau~$ samples per second. The carrier amplitude $a$ determines the transmitted power $P_t = 10 \ \mathrm{log}_{10}((a/2)^2/10^{-3})~\mathrm{dBm}$, assuming both symbols are equiprobable. 
We strongly oversample $s_{\mathrm{TX}}(t)$ (at $1.1\times 10^{12}~\mathrm{Hz} = 18.5f_0$) to avoid sampling artefacts.

Second, we aim to determine the received signal $s_{\mathrm{RX}}(t)$. To that end, we begin by Fourier transforming $s_{\mathrm{TX}}(t)$, yielding $S_{\mathrm{TX}}(f)$. Then, we multiply $S_{\mathrm{TX}}(f)$ with the channel $S_{12}(f)$ to obtain $S_{\mathrm{RX}}(f)=S_{12}(f)S_{\mathrm{TX}}(f)$. The received signal $s_{\mathrm{RX}}(t)$ is subsequently obtained as the real part of the inverse Fourier transform of $S_{\mathrm{RX}}(f)$. This procedure is equivalent to convoluting $s_{\mathrm{TX}}(t)$ with the CIR $h_{12}(t)$ in the time domain. Moreover, we add Gaussian noise $n(t)$ with zero mean and standard deviation $\chi$ to the received signal. This emulates thermal noise at the receiver that is independent of the received signal; the noise power is consequently $P_n = 10 \ \mathrm{log}_{10}(\chi^2/10^{-3})~\mathrm{dBm}$.

\begin{equation}
s_{\mathrm{RX}}(t) = s_{\mathrm{TX}}(t) \ast h_{12}(t) + n(t).
\label{eq:s_RX}
\end{equation}

Third, the signal $s_{\mathrm{RX}}(t)$ at the receiver is sampled and quantized to obtain the received signal stream $\mathbf{b_r}$. Specifically, we begin by dividing $s_{\mathrm{RX}}(t)$ into intervals of length $\tau$ which are shifted with respect to the transmitted signal by the propagation delay $\Delta_{pd}$ associated with the first arriving pulse. Then, we implement energy detection by integrating $s_{\mathrm{RX}}^2(t)$ over each interval, yielding a $1\times N$ vector of detected energies $\mathbf{d}$.

\begin{equation}
\mathbf{d}[i] = \int_{(i-1)\tau+\Delta_{pd}}^{i \tau+\Delta_{pd}} s_{\mathrm{RX}}^2(t).
\label{eq:d}
\end{equation}

\noindent Next, we compare each entry of $\mathbf{d}$ to a threshold value $d_ {\mathrm{thresh}}$ and assign `1' if the value is above the threshold, and `0' otherwise. This finally yields the decoded data stream $\mathbf{b_r}$. We identify the threshold as the energy value that minimizes the BER, evaluated over a long series of random symbols.

\begin{equation}
\mathbf{b_r}[i] = \begin{cases}
0 &\text{if $\mathbf{d}[i] \leq d_\mathrm{thresh}$.}\\
1 &\text{if $\mathbf{d}[i] > d_\mathrm{thresh}$.}
\end{cases}
\label{eq:b_r}
\end{equation}

In practice, if $N$ is large, we perform the above-detailed OOK simulation procedure for shorter subsections of $\mathbf{b_t}$, one after the other. In order to not distort ISI effects through this procedural detail, each subsection begins with the last three bits of the previous subsection.

In order to evaluate the BER, we use a random binary data stream with $N=2.28\times 10^5$ symbols of equal probability, i.e., $\langle b(t) \rangle _t = 0.5$. 

In order to transfer an $A\times B$ full-color image via the wireless channel, we first flatten the 3D matrix representing it (two spatial dimensions, one dimensions for the three RGB color components), yielding a $1\times 3AB$ vector. Then, we replace each 8-bit entry of this vector with its corresponding binary representation, yielding a $1\times 24AB$ vector. This bit stream is then transferred via the wireless channel following the previously outlined procedure. The received signal is ultimately assembled based on the received and decoded data stream of equal length by performing the procedure outlined in this paragraph in reverse order.

\section*{Acknowledgements}
S.A. is supported by the European Commission through the H2020 FET-OPEN program under grants No. 736876 and No. 863337.

\section*{Data Availability}

The data that support the findings of this study are available from the corresponding author upon reasonable request.

\section*{Code Availability}

Code that supports the findings of this study is available from the corresponding author upon reasonable request.

\section*{Author Contributions}

P.d.H. conceived the project. 
M.F.I. designed the meta-atoms and optimized the RIS configurations, with input from S.A. and P.d.H.
P.d.H. performed the OOK simulations, with input from M.F.I. and S.A.
P.d.H. wrote the manuscript. 
All authors contributed with thorough discussions and reviewed the manuscript.


\providecommand{\noopsort}[1]{}\providecommand{\singleletter}[1]{#1}%

\end{document}